\documentclass[aps,preprint,prb,showpacs,amsmath,superscriptaddress,onecolumn,floatfix,nobalancelastpage]{revtex4-1}

\usepackage[nottoc,numbib]{tocbibind}

\usepackage[colorlinks=true, citecolor=black, linkcolor=black, urlcolor=black]{hyperref}

\usepackage[all]{hypcap}
\usepackage{amsmath}
\usepackage{enumerate,bm}
\usepackage{graphicx}
\usepackage{grffile} % to handle dots in graphics filnames
\usepackage{color}
\usepackage{esint}
\usepackage{textpos}
\usepackage[usenames,dvipsnames]{xcolor}
\usepackage{subfigure}
\usepackage{flushend}
\usepackage{tabularx}
\usepackage{amssymb}
\usepackage{float}
\usepackage{subfigure}
\usepackage{ulem} %striketrhough

\usepackage{setspace}

\begin{document}
% \linenumbers
%%%%%%%%%%%%%%%%%%%%%%%%%%%%%
%%%%%%%%%          AUTHORS              %%%%%%
%%%%%%%%%%%%%%%%%%%%%%%%%%%%%

%Nature Physics Article

\author{Umberto~De~Giovannini}
\email{umberto.degiovannini@gmail.com}
\affiliation{Nano-Bio Spectroscopy Group and ETSF, Universidad del Pa\'is Vasco, CFM CSIC-UPV/EHU, 20018 San Sebasti\'an, Spain}

\author{Hannes~H\"ubener}
\email{hannes.huebener@gmail.com}
\affiliation{Nano-Bio Spectroscopy Group and ETSF, Universidad del Pa\'is Vasco, CFM CSIC-UPV/EHU, 20018 San Sebasti\'an, Spain}

\author{Angel~Rubio}
\email{angel.rubio@mpsd.mpg.de}
\affiliation{Nano-Bio Spectroscopy Group and ETSF, Universidad del Pa\'is Vasco, CFM CSIC-UPV/EHU, 20018 San Sebasti\'an, Spain}
\affiliation{Max Planck Institute for the Structure and Dynamics of Matter and Center for Free Electron Laser Science, 22761 Hamburg, Germany}

\title{Monitoring in real time the photon-dressing and undressing of quasiparticles from first principles time-resolved photoelectron spectroscopy}

\date{\today}
\begin{abstract}
Optical pumping of solids creates a non-equilibrium electronic structure where electrons and photons combine to form quasiparticles of dressed electronic states. The resulting shift of electronic levels is known as the optical Stark effect, visible as a red shift in the optical spectrum. Here we show that in a pump-probe setup we can uniquely define a non-equilibrium quasiparticle bandstructure that can be directly measurable with photo-electron spectroscopy. The dynamical photon-dressing (and undressing)  of the many-body electronic states can be monitored by pump-probe time and angular resolved photoelectron spectroscopy (tr-ARPES) as the photon-dressed bandstructure evolves in time depending on the pump-probe pulse overlap.  The computed tr-ARPES spectrum agrees perfectly with the quasi-energy spectrum of Floquet theory at maximum overlap and goes to the the equilibrium bandstructure as the pump-probe overlap goes to zero. Additionally, we show how this time-dependent non-equilibrium quasiparticle structure can be understood to be the bandstructure underlying the optical Stark effect. The extension to spin-resolved PES can be used to predict asymmetric dichroic response linked to the valley selective optical excitations in monolayer transition metal dichalcogenides (TMDs).
\end{abstract}

\maketitle

\section{Introduction}
Laser driving of solids is believed to give rise to a steady-state non-equilibrium phase of the electronic structure where quasiparticles of combined photons and electrons emerge with a distinct band structure, so called Floquet-Bloch bands.\cite{Faisal:1997fq} At off-resonant driving photon energy is not absorbed by electronic states but they dress the electronic structure via virtual photon processes, leading to distinctive replica bands at multiples of the driving energy. These photon dressed electronic states can have dramatically different properties from those of their equilibrium counter parts. In particular the possibility to induce topological phases on an ultrafast time-scale has received much attention recently, following the proposal of a Floquet-topological insulator.\cite{Lindner:2011ip} In such a system the coupling to a monochromatic light field can be used to induce, in an ordinary equilibrium semiconductor, non-trivial topological order leading to the emergence of topologically protected edge states. Similar mechanisms have been proposed to engineer the topology in 2D Dirac materials.\cite{Claassen:2016ti,Sentef:2015jp,Morell:2012hh,Kitagawa:2011fj} Early on it was realised that the Floquet mechanism can also be used to create solid state versions of the elusive Majorana Fermions,\cite{Jiang:2011cw,Li:2014dw,Benito:2014bd} that have still not been directly observed.  With the discovery of bulk Dirac semimetals\cite{Liu:2014bf,Ali:2014kc} the scope for Floquet materials has considerably widened, in particular the creation of non-equilibrium Floquet-Weyl Fermions has been discussed.\cite{Hubener:2016um,Zou:2016do,Wang:2016dw} Other recent Floquet-topological phenomena include nodal rings,\cite{Yan:2016ee} type-II Floquet-Weyl semimetals\cite{Chan:2016uo}, carbon nanotubes\cite{Hsu:2006bg} and soundwaves\cite{Fleury:2016cr} among others.   

While this shows how the number of theoretical proposals for Floquet-topological phenomena is currently growing at a fast pace, experimental evidence for the existence and measurability of the underlying Floquet-bandstructure is somewhat lacking. Only two recent experiments unambiguously found direct proof of Floquet states on the surface of a topological insulator, via tr-ARPES spectroscopy.\cite{Wang:2013fe,Mahmood:2016bu} Yet, it is commonly assumed that such bands appear in any kind of material and particularly in the bulk phase. Tr-ARPES is the method of choice for monitoring many groundstate and excitation phenomena in solids, giving access to the time-dependent development of spatial, momentum and spin degrees of freedom of single energy levels. In this work we use first-principles computational techniques to show under which conditions time-resolved pump-probe ARPES can measure Floquet-bandstructures in a semiconductor and provide a protocol for observing the creation of the emerging bands in real-time. The pump pulse creates new quasiparticle states that are subsequently probed by a second pulse and depending on the time delay between the two pulses the creation and destruction of the quasiparticles is observed. This non-equilibrium bandstructure is not only underlying the Floquet-topological phenomena but is also occurring in more conventional pump-probe experiments. The formation of the new quasiparticle levels leads to a shift in the absorption energy of the pumped system, known as the optical or dynamical Stark effect.\cite{Autler:1955gb} In a recent work, optical pump-probe spectroscopy was used to demonstrate that the Stark effect can be used in TMDs to confirm selective excitations of the electronic structure in certain valleys of the conduction band.\cite{Sie:2015hn} Here, we show how this effect emerges from the Floquet-bandstructure and how the spin polarization of the valence bands results in a spin-selective interaction of the dressed quasiparticle states.

We stress that in this work we use two different theoretical approaches based on time-dependent density functional theory (TDDFT).\cite{runge_density-functional_1984} We directly simulate the ARPES measurement of a pumped system by computing the time-evolution of the electronic density under presence of the two pulses and analyse the energy and momentum distribution of the ionized photo-electrons.\cite{DeGiovannini:2012hy,Wopperer:2016ur,UDG2016} On the other hand we perform Floquet analysis of the time-evolution of the pumped system, without any reference to a probe pulse, but imposing periodicity in time.\cite{Hubener:2016um} The striking finding is that both methods give bandstructures that agree nearly perfectly with each other. In our simulation we do not include coupling to lattice degrees of freedom, which can also affect the non-equilibrium bandstructure via scattering with phonons. However, here we are considering a monolayer TMD which has a direct bandgap, so that for low excitation energies there is no significant electron-phonon scattering through the Brillouin zone. While other dissipation effect can be at play here, what we show is only the underlying electronic eigenstates of the driven system and how they emerge from the equilibrium bandstructure. A complete description of optical pumping, as suggested here, requires taking into account electron-hole interactions, which we have omitted. For the sake of clarity, we argue that the effect we discuss is generic and that it applies similarly to excitonic states.

%here methods for PRL

\section{Dressing and undressing of quasiparticles}
Transition metal dichalcogenides have emerged as a platform to study many new phenomena of excited electronic structures. Their 2D stacking structure makes them particularly suitable to study monolayers where the broken inversion symmetry results in a well defined spin texture of the valence and conduction band edge. Together with a multivalley structure of the conduction bands this property makes them particularly interesting for possible spin- and valleytronics applications.\cite{Xu:2014go} Furthermore, the  band edges have orbital-character with well defined angular momenta so that there is a pronounced valley-selective dichroism. Recently it has been shown that also in the bulk phase most of these properties persist with each layer of the material carrying a well defined valley-dependent spin polarization.\cite{Zhang:2014cl,Riley:2014bw,Bertoni:2016ur} Here we consider monolayer WSe$_2$ to utilize these properties to create a dressed-quasiparticle structure with a distinct valley and spin dependence. 

We simulate the pump-probe photoemission process and angle resolved measurement, as depicted in Fig.~1a, where the pump is a circularly polarized monochromatic pulse that is long enough to drive the electronic structure into a non-equilibrium but stationary state, which usually is a few cycles of a field with $\sim 1$eV. Then the probe is applied with typically a higher photon energy that is large enough to ionize electrons from the sample, here we use a UV pulse with 40eV. The fast oscillating probe field has an envelop of $\sim85$fs, so that during the probing process the driving field completes several tenth of cycles. Thus the measurement process effectively averages the oscillating non-equilibrium state over many of its cycles. The atomic structure of monolayer WSe$_2$, is depicted in Fig.~1b, along with the Brillouin zone and the path across the $K$ point that we are considering here. The observed photo-electron spectrum depends strongly on the overlap of the pump and probe pulses as shown in Fig.~1d. When the two pulses do not overlap one only measures the equilibrium bandstructure of WSe$_2$, but in case of overlapping pulses extra features occur in the spectrum. The non-equilibrium state of the driven monolayer consists of quasiparticles that are a combination of electrons from the material and photons from the driving field, so called photon-dressed electronic states. Thus the photon-dressed features rely on the fact that the pump and probe pulses are overlapping, while most commonly tr-ARPES measurements are used to probe the electron dynamics following an excitation and to monitor its decay dynamics. 

The dressing of the electronic bands leads to replicas of the equilibrium bands shifted by the photon energy. In a simple non-interacting picture this would results in a bandstructure with copies of all bands shifted from the equilibrium position by the pump energy. However, under interacting conditions, when the replica of a band is shifted such that it gets close to another equilibrium band the two can hybridize as schematically shown in Fig.~1c. In particular, when the pump frequency has the energy of the bandgap, then replicas of the valence band top are moved to the conduction band bottom and vice versa. These dressed bands are directly observable in the ARPES spectra of Fig.~1d. When the temporal overlap between pump and probe pulse is modified, we can observe how the dressed bands collapse to the equilibrium bands. This means it is possible to directly observe the creation and destruction of the photon-electron quasiparticle by tuning the pump-probe overlap: the real-time dressing and undressing of an electron. 

\begin{figure}
    \centering
     \resizebox{0.8\textwidth}{!}{\includegraphics{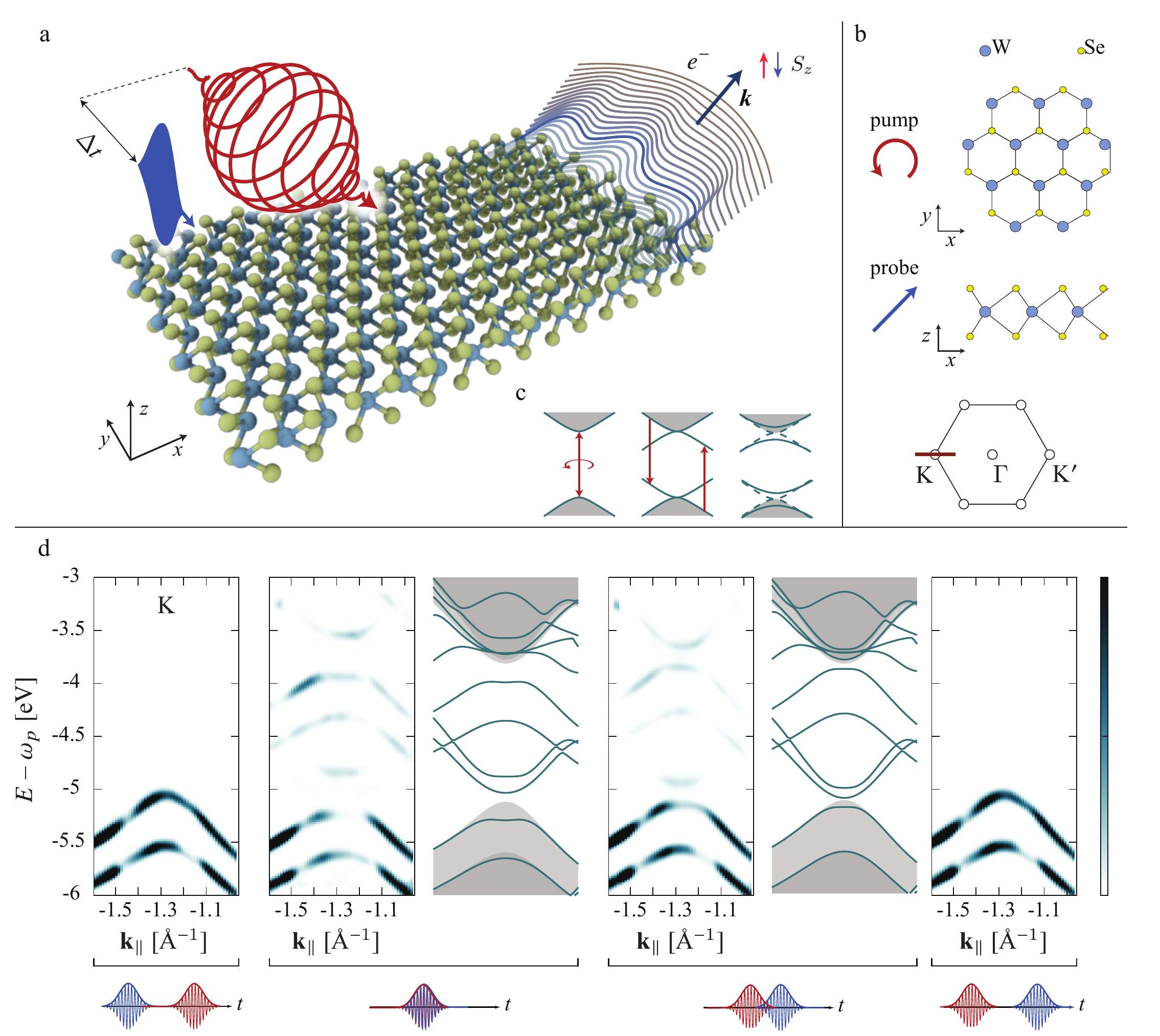}}
    \caption{{\bf Dressing and undressing.} \textbf{a} Schematic of the computational pump-probe ARPES set-up. The electronic structure of monolayer WSe$_2$ is driven out of equilibrium by a pump pulse (red) and after some time delay $\Delta t$ a second pulse (blue) is used to excite the electronic structure such that electrons detach from the sample with energies depending on the crystal momentum $\mathbf{k}$ reflecting the non-equilibrium bandstructure. \textbf{b} The WSe$_2$ monolayer is aligned with the $x$-$y$ plane as is the polarization plane of the circularly polarized pump pulse (red). The probe pulse (blue) is linearly polarized in the $x$-$z$ direction and the photo-electron spectrum is computed on a short path in the Brillouin zone crossing K in the K-$\Gamma$ direction.\textbf{c} Driving a semiconductor at resonance with its band gap leads to the formation of photo-dressed states at $\pm$ the driving energy thus creating replicas of the valence band top and conduction band minimum that are respectively degenerate with the conduction band minimum and valence band top, leading to a hybridisation that opens the overall gap. \textbf{d} Depending on the delay $\Delta t$ between the pump and the probe pulses different non-equilibrium bandstructures are observed. When the system is probed before the pump is applied the photo-electron spectrum gives the equilibrium valence bands. When the probe is overlapping completely with the pump pulse the photon-dressed quasiparticle bands are visible in the photo-electron spectrum corresponding to the bandstructure obtained with Floquet-theory. The quasiparticle picture for photo-dressed electrons holds even when the the pump is slowly switched off, corresponding to partial overlap of the two pulses. Also in this case the observed photo-electron spectrum corresponds to a bandstructure obtained with Floquet theory, but for the pump amplitude at the maximum of the probe pulse. When the system is probed after the pump is switched off, all photon-dressed quasiparticle features have disappeared. Thus td-ARPES allows the observation of the formation and collapse of the photo-dressed quasiparticle.}
    % = 316 words on Monday 08/08/2016 10:48 
\end{figure}

\section{Floquet theory}
The observed quasiparticle structure can be understood in terms of Floquet theory,\cite{Sambe:1973hi,Hone:1997db} where a stationary state is expanded into a basis of Fourier components of multiples of the photon frequency $\Omega$: $|\psi(t)\rangle=\sum_{m}\exp(-i(\epsilon+m\Omega)t)|u_m\rangle$, where $\epsilon$ is the equilibrium energy of the state. With this ansatz the time-dependent Schr\"odinger equation becomes an eigenvalue problem $\sum_{n}\mathcal{H}^{mn} |u_n\rangle = \epsilon |u_m\rangle$ of the static Floquet Hamiltonian $\mathcal{H}^{mn}  = \frac{\Omega}{2 \pi}\int_{2\pi/\Omega} dt e^{i(m-n)\Omega t} H(t) + \delta_{mn}m\Omega$. The eigenstates of this Hamiltonian span a Hilbert space with the dimension of the original electronic Hilbert space times the multiphoton dimension. The contribution of the latter is in principle infinite, but here can be truncated. The spectrum of this Hamiltonian gives the bandstructure of the dressed quasiparticles. We compute the Floquet bandstructure using Floquet-TDDFT\cite{Hubener:2016um} where the time-dependent Hamiltonian of TDDFT is used in the Floquet analysis described above. We point out that we do not perform a high-frequency expansion, but rather restrict the Floquet Hamiltonian to contain only $\pm\Omega$ terms. 

In Fig.~1d are also shown, side-by-side with the calculated ARPES spectra, the Floquet bands corresponding to the pump parameters used in the ARPES simulation. The dressed bands observed in the ARPES calculation are perfectly reproduced. We note that only certain bands hybridise while others make clean crossings. This is due to the spin polarization of the states and will be discussed below in more detail. Here we would like to point out the core finding of the work: The dressed states can be interpreted as Floquet-sidebands and the measurement process in ARPES with a finite-width probe pulse can in turn be interpreted as performing a time average of the non-equilibrium oscillating electronic structure corresponding to the time integrals in the Floquet analysis . In particular, the undressing process can be described in terms of a Floquet bandstructure with an effective constant amplitude, despite the fact that the amplitude is actually varying over the time during which the probe is applied. The undressing occurs while the the pump is being switched off and thus the amplitude is actually changing while the system is probed by the finite probe pulse. This means that the system is not strictly a time-periodic stationary state. However, the tr-ARPES bandstructure computed under this condition perfectly agrees with the Floquet bandstructure that by definition assumes a constant pumping amplitude, and thus is strictly speaking not applicable to this situation, because it represents a transient state. In the case of Fig.~1d the effective amplitude that reproduces the switching-off bandstructure corresponds to the amplitude of the shaped pump pulse at the time where the center of the probe pulse is located. Again we point out that these two matching bandstructures are obtained independently with two different computational approaches. This underlines how the Floquet expansion is not only the correct way to describe the time-averaging inherent in a probe measurement, but that it is also applicable even when the system is not strictly stationary. Clearly, this is only the case when the transient components of the oscillation are small and therefore the error committed in using Floquet analysis is small. For faster changing non-equilibrium states this will certainly break down. 

The striking agreement between Floquet theory and ARPES calculation allows to further explore the parameter space of the probe pulse. Figs.~2(a-c) show the dressed bands for the same pump energy but with different amplitudes from those in Fig.~1d. The splitting of the bands due to the hybridisation can be seen to strongly depend on the strength of the pump. While the amplitude of the pump controls the splitting of the degenerate photon-dressed levels, the pump energy controls the shift of the bands, as noted before. As shown in Fig.~2d, that means that for a pump energy smaller than the bandgap the photon-dressed states are observable as additional bands inside the band gap, but the interaction with the original equilibrium bands is much weaker, because they are not at degenerate energies. The Floquet bandstructure for this case shows that there is an avoided crossing behaviour between different photon-dressed bands. These features are much more pronounced when the driving energy is larger than the bandgap, so that the dressed states are moved well into the valence and conduction band. We then observe multiple avoided crossings and in general a much richer spectrum of complex intersecting bands. 

\begin{figure}
    \centering
     \resizebox{0.9\textwidth}{!}{\includegraphics{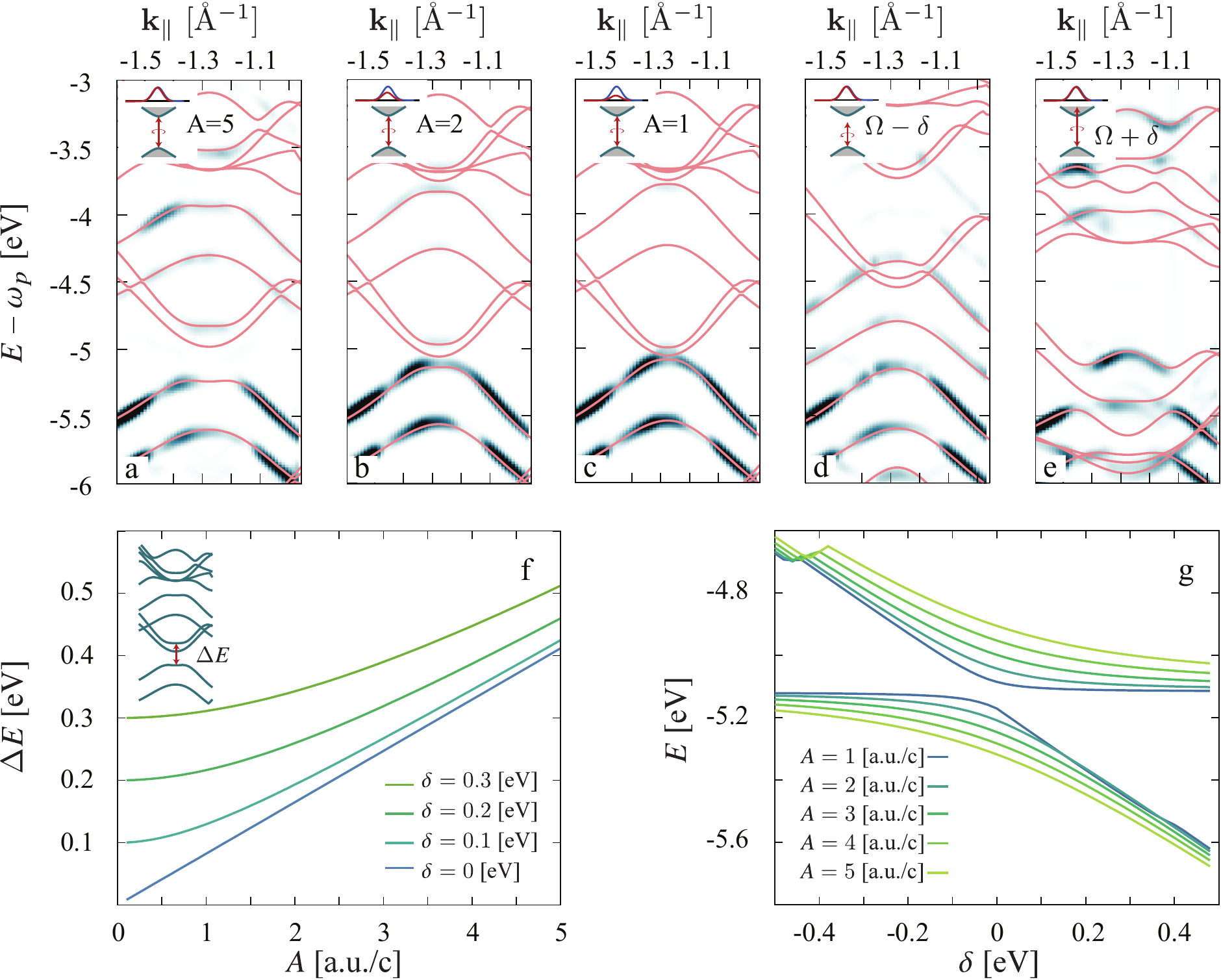}}
    \caption{{\bf Dependence on laser parameters.} The amplitude and energy of the pump pulse determines the photon-dressed quasiparticle bandstructure in accordance with results from Floquet-theory (pink lines). \textbf{a-c} When the energy is in resonance with the bandgap the photon-dressed states hybridise with the equilibrium valence and conduction bands, where the amount of the hybridisation depends on the amplitude of the pump pulse. \textbf{d} When the energy of the pulse is smaller than the bandgap only replicas of the valence bands are observed. \textbf{e} Pumping with energy larger than the bandgap leads to a complex quasiparticle bandstructure with multiple interactions of photon-dressed bands with equilibrium bands. \textbf{f} Using the bandstructures resulting from Floquet theory the scaling of the hybridization gap (see main text for a discussion) between the photo-dressed conduction band minimum and the valence band top (inset) is shown as a function of the pump laser amplitude for different detuning energies, behaving quadratically as expected from the ac-Stark effect. \textbf{g} The hybridisation gap of (\textbf{f}) can be seen as an avoided crossing of the two bands when scanning the pump energy through the resonance with the bandgap.}
    % = 180 words Monday 08/08/2016 11:20
\end{figure}

\section{scaling}
Now, using results from Floquet theory we can systematically elucidate this behaviour. Fig.~2f shows the energy shift of the top valence band at $K$ as a function of the amplitude for different pump energies. We observe that there is qualitative difference between the resonant and those pumps that are detuned from the band gap. While in the resonant case the shift grows linearly with the amplitude, detuning initially leads to a quadratic onset and only for large amplitudes a linear behaviour is observed. This different scaling behaviour can be rationalised by considering a simple driven two level model system with the Hamiltonian 
\begin{equation}
H(t)=\frac{\epsilon}{2}(|c\rangle\langle c| -|v\rangle\langle v|) + (M|v\rangle\langle c| + c.g. ) A \cos(\Omega t) 
\end{equation}
where, $\epsilon$ is the energy gap, $M=\langle v| r |c\rangle$ is the dipole matrix element between the two states and $A$ the amplitude of the driving field. Constructing the Floquet Hamiltonian for this system while considering only one-photon interactions, further restricting it to one-photon absorption (see Supplementary Information) and diagonalising it, gives the shift of the levels as a function of amplitude and frequency of the pump laser as $\Delta E = \sqrt{4 A^2 M^2 + (\Omega-\epsilon)^2}$. If $\Omega$ is in resonance with the gap, $\Omega\rightarrow \epsilon$, this shift simplifies to $\Delta E_{\rm res} = 2 A M$, i.e. it grows linearly as a function of the field intensity. In case of detuning, $\Omega\rightarrow \epsilon + \delta$,   the energy shift is $\Delta E_{\rm det} = \sqrt{4 A^2 M^2 + \delta^2}$. Expanding this for small amplitudes gives $\Delta E_{\rm det} \approx \delta + \frac{2 A^2 M^2}{\delta}$. Thus for off-resonant driving and $AM<<\delta$ the energy shift grows quadratically with the amplitude, as observed in Fig.~2f.

\section{Optical Stark effect}
In fact, a quadratic dependence on the amplitude of a light-induced energy shift is the well known behaviour the optical Stark effect\cite{Sussman:2011hl} which emerges thus naturally from Floquet theory. This highlights the fact that the dressed bands observed in the ARPES calculation are the underlying electronic structure of the optical Stark effect. In particular, in a recent experiment Sie~et.~al.\cite{Sie:2015hn} demonstrate a valley selective optical Stark effect in a transition metal dichalcogenide with an all optical pump-probe measurement. Here, we can discuss this experiment in terms of the underlying electronic quasiparticle process, as it is accessible through ARPES and provide a detailed picture of how the optical Stark effect occurs on the level of the electronic structure.

Monolayer WSe$_2$ is known to have circular pump dichroism,\cite{Cao:2012eu} originating from a well defined angular momentum character of electronic orbitals at the $K$ and $K'$ points in the Brillouin zone.\cite{Berghauser:2014ce} Thus, a dipole selection rule determines that for a given circularly polarized pump the electronic density at only one of the two $K$ valleys is excited. Fig.~3 shows how pumping with left-handed circularly polarized light results in the formation of photon-dressed bands only in the $K$-valley and thus in a valley selective optical Stark effect. The hybridization of the bands leads to a down shift of the valence band and in an up shift of the conduction band and thus in an overall opening of the equilibrium band gap at the $K$ valley. Also, there is a spin dependence of this process that would clearly manifest in a spin resolved td-ARPES measurement. When electronic excitation is prevented by dipole matrix elements, only the valence bands appear as photon-dressed replicas in the photoemission spectrum while the absence of electronic population in the conduction bands implies that there are no electronic bands available to be dressed, as seen in Fig. 3. Since the Stark effect is a result of the hybridisation between photon-dressed valence and conduction bands with the respective undressed bands, the absence of population in the conduction bands in the $K'$-valley precluding any hybridization thus implies that the Stark effect does not occur. In an optical absorption experiment, this selective gap opening is observable as a shift of the absorption peak\cite{Sie:2015hn} of circularly polarized light with the same helicity as the pump, while a probe  with the opposite helicity only interacts with the unshifted bandstructure of the $K$ valley that is unaffected by the circular pump.

Our calculation also elucidates the role of spin polarization in the observable Stark effect. The broken inversion symmetry of the monolayer with respect to the symmetric bulk phase leads to a spin-orbit splitting of the valence and conduction bands with a nearly perfect spin polarization of the two bands at $K$.\cite{Riley:2014bw} We have noted before that in the Floquet bandstructures in Figs.~1 and 2 only certain bands appear to be hybridising. From Fig.~3 this effect becomes clear and can be identified to originate from the spin polarization of the equilibrium bands: The photon-dressed states inherit the spin polarization from the underlying equilibrium bandstructure and therefore the hybridisation is spin selective, i.e. only bands with the same spin-polarization will hybridise, see Fig.~3e. In particular, this means that a pump energy equal to the band gap does not actually move the photon-dressed states into perfect resonance because the states at the band edges have opposite spin polarisation. Instead to achieve exact resonant condition the required energy is equal to the difference between a conduction and valence band with the same spin polarization, as has been used in Figs.~2(f-g). 

\begin{figure}
    \centering
     \resizebox{0.9\textwidth}{!}{\includegraphics{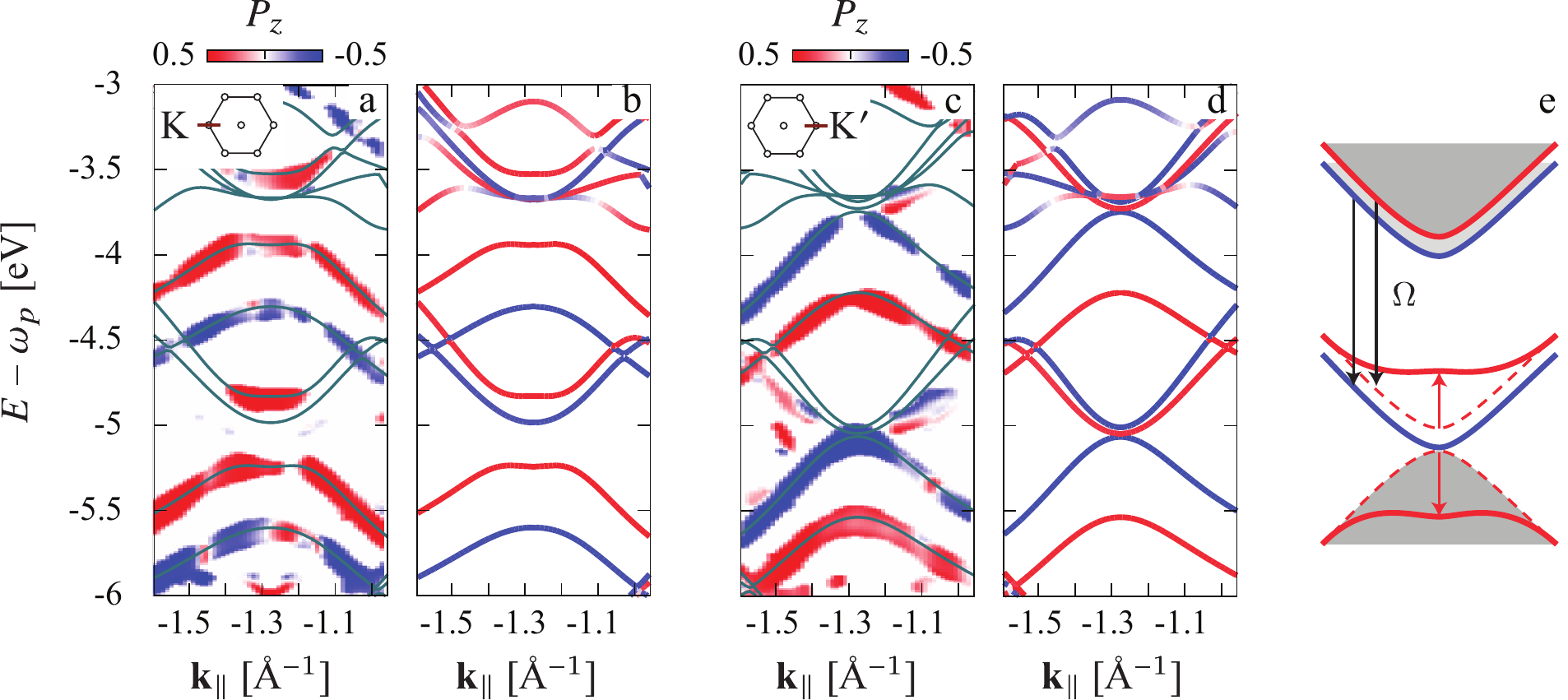}}
    \caption{{\bf Dichroic optical Stark effect.} Pump-probe ARPES together with Floquet theory reveals the role of spin-polarization in the quasiparticle structure underlying the optical Stark effect in WSe$_2$. \textbf{a}-\textbf{b} The spin-polarization of the photo-dressed states is inherited from the equilibrium bands. Since the valence and conduction bands at K are strongly polarized in the $S_z$ component, hybridisation occurs only between certain bands. \textbf{c}-\textbf{d} The pump dichroism of WSe$_2$ results in the absence of the Stark effect at the time-reversed K$^\prime$ point. Due to the pump helicity electronic excitations in this valley are forbidden by dipole selection and hence no photon-dressing of conduction bands occurs and only the photon-dressed replicas of the valence bands are observed.
    \textbf{e} The photo-dressed states hybridize with bands that are close in energy but only within the same spin polarization subspace.}
    %= 127 words on Monday 08/08/2016 11:53
\end{figure}

\section{Discussion}
%In this work we have used two different first-principles computational techniques to compute the non-equilibrium electronic structure and its real-time measurement process under pump-probe condition of the semiconductor WSe$_2$. Our \textit{ab initio} method for the photo-electron spectrum is a direct simulation not only of the electronic excitation due to the pump pulse but also of the photo-emission process following stimulation by a probe pulse with energy above the ionization threshold as well as the subsequent momentum analysis of the free electrons as occurring in an ARPES experiment. While this method makes no assumption on the non-equilibrium state of the electronic structure, our second method, Floquet-TDDFT, requires that the system is in a stationary state that is periodic in time. Remarkably, both methods give the same non-equilibrium band structure.

%We have shown that by controlling the overlap between pump and probe pulses in a td-ARPES set-up one can directly observe the dynamical formation of the photon-dressed quasiparticle bandstructures. Such a measurement gives the bandstructure underlying the optical Stark effect and can be analysed in terms of the Floquet bandstructures. The Floquet analysis of a simple two level system can be shown to result in the well known expression of the ac-Stark shift from perturbation theory. We systematically show the scaling of the optical Stark effect with the pump field parameters also beyond the perturbative regime.  

The photon dressing of electrons to form non-equilibrium quasiparticles is the fundamental process underlying the vast amount of theoretical proposals in the rapidly developing field of Floquet-topological materials. Here we describe the conditions for the direct observation of such bands in a semiconducting monolayer TMD. We have discussed the role of the photon dressed states in terms of the optical Stark effect, but the Floquet quasiparticle structure of monolayer TMDs also hosts topological properties. In equilibrium, most monolayer TMDs are ordinary semiconductors but it has been recently proposed\cite{Claassen:2016ti} that they can be driven by monochromatic light to form a Floquet-topological insulator. The energy of the protected edge states of such a non-equilibrium phase cross the hybridisation gap between the photon-dressed state and an equilibrium band and including an edge geometry in our calculation would show also this band. There is thus a direct connection between the bandstructure presented here and the Floquet topological phase. Our work shows by first principles calculations that the measurement of the non-equilibrium bands with ARPES is intimately linked to the Floquet quasiparticle spectrum. Indeed, by using two completely independent computational methods we systematically obtain the same non-equilibrium bandstructures. We believe that this work paves the way to understanding the measurement of dynamical photon dressing phenomena in a large variety of materials in non-equilibrium phases.  

\section{Acknowledgements}
We are grateful for illuminating discussions with M. Sentef and R. Ernstorfer. We acknowledge financial support from the European Research Council (ERC-2015-AdG-694097), Spanish grant (FIS2013-46159-C3-1-P), Grupos Consolidados (IT578-13), AFOSR Grant No. FA2386-15-1-0006 AOARD 144088, and European Union’s Horizon 2020 Research and Innovation program under Grant Agreements no. 676580 (NOMAD) and 646259 (MOSTOPHOS). H.H. acknowledges support from the People Programme (Marie Curie Actions) of the European Union's Seventh Framework Programme FP7- PEOPLE-2013-IEF project No. 622934. 

\section{Methods}
The evolution of the electronic structure under the effect of external fields was computed by propagating the Kohn-Sham (KS) equations in real-space and real time within TDDFT as implemented in the Octopus code~\cite{Strubbe:2015iz}. 
We solved the KS equations in the local density approximation (LDA)~\cite{Perdew:1981dv} with semi-periodic boundary conditions. We used a simulation box of 120~$a_0$ along the non-periodic dimension and the primitive cell on the periodic dimensions with a grid spacing of 0.4~$a_0$. 
We modeled WSe$_2$ with a lattice parameter of 6.202~$a_0$ and by sampling the Brillouin zone with a 12$\times$12 k-point grid.  
Time and spin-resolved ARPES was calculated by recording the flux of the photoelectron current over a surface placed 30~$a_0$ away from the system with the t-SURFFP method~\cite{UDG2016}.
All calculations were performed using fully relativistic HGH pseudopotentials~\cite{Hartwigsen:1998dk}.

The Floquet bandstructures of the driven system were computed using Floquet-TDDFT~\cite{Hubener:2016um} in the non-interacting particle approximation. The spin-resolved Floquet bands where obtained by first expanding the Floquet-eigenstates $\psi^{\rm F}$ in the basis of equilibrium Kohn-Sham states $\psi^{\rm KS}$:
\begin{equation}\label{eq:expansion}
 |\psi_n^{\rm F}\rangle = \sum_i c_{ni}|\psi_i^{\rm KS}\rangle.
\end{equation}
Using the spin-density functional theory extension to DFT,\cite{vonBarth:1972eq} each of these Kohn-Sham states is a two-spinor and has well defined spin projections. Here, we consider the $S_z$ component and compute the spin content of the Floquet states as
\begin{equation}
  \langle S_z \rangle_n^{\rm F} = \sum_i c_{ni} \langle S_z \rangle_i^{\rm KS} .
\end{equation}
where $c$ are the coefficients of Eq.~(\ref{eq:expansion}). Thus the Floquet bands inherit the spin from the equilibrium bands.

In this work we do not include dissipation effects, but they are straightforward to include in time-resolved calculation by coupling the evolution of the electrons to the lattice degrees of freedom via the Ehrenfest theorem.\cite{Andrade:2009ga}

\bibliography{td-ARPES}
\end{document}